\documentstyle[aps,floats,psfig,cite]{revtex}
\begin{document}
%
%%%%%%%%%%%%%%%%%%%% TITLE PAGE %%%%%%%%%%%%%%%%%%%%%%%%%%%%%%%%%%%%%%%%%%%%%%
%
\title{Search for Second Generation Leptoquark Pairs Decaying to $\mu\nu + jets$
 in $p\overline{p}$  Collisions at $\sqrt s$ = 1.8 TeV}
%
%
% LIST_OF_AUTHORS                4/20/99            
%
\author{                                                                      
%% names begin here                                                           
B.~Abbott,$^{45}$                                                             
M.~Abolins,$^{42}$                                                            
V.~Abramov,$^{18}$                                                            
B.S.~Acharya,$^{11}$                                                          
I.~Adam,$^{44}$                                                               
D.L.~Adams,$^{54}$                                                            
M.~Adams,$^{28}$                                                              
S.~Ahn,$^{27}$                                                                
V.~Akimov,$^{16}$                                                             
G.A.~Alves,$^{2}$
% ...
\\                                                                            
\vskip 0.30cm                                                                 
\centerline{(D\O\ Collaboration)}                                             
\vskip 0.30cm                                                                 
}                                                                             
\address{                                                                     
\centerline{$^{1}$Universidad de Buenos Aires, Buenos Aires, Argentina}       
\centerline{$^{2}$LAFEX, Centro Brasileiro de Pesquisas F{\'\i}sicas,         
                  Rio de Janeiro, Brazil}                                     
\centerline{$^{3}$Universidade do Estado do Rio de Janeiro,                   
                  Rio de Janeiro, Brazil}                                     
\centerline{$^{4}$Institute of High Energy Physics, Beijing,                  
                  People's Republic of China}                                 
\centerline{$^{5}$Universidad de los Andes, Bogot\'{a}, Colombia}             
\centerline{$^{6}$Universidad San Francisco de Quito, Quito, Ecuador}         
\centerline{$^{7}$Institut des Sciences Nucl\'eaires, IN2P3-CNRS,             
                  Universite de Grenoble 1, Grenoble, France}                 
\centerline{$^{8}$DAPNIA/Service de Physique des Particules, CEA, Saclay,     
                  France}                                                     
\centerline{$^{9}$Panjab University, Chandigarh, India}                       
\centerline{$^{10}$Delhi University, Delhi, India}                            
\centerline{$^{11}$Tata Institute of Fundamental Research, Mumbai, India}     
\centerline{$^{12}$Kyungsung University, Pusan, Korea}                        
\centerline{$^{13}$Seoul National University, Seoul, Korea}                   
\centerline{$^{14}$CINVESTAV, Mexico City, Mexico}                            
\centerline{$^{15}$Institute of Nuclear Physics, Krak\'ow, Poland}            
\centerline{$^{16}$Institute for Theoretical and Experimental Physics,        
                   Moscow, Russia}                                            
\centerline{$^{17}$Moscow State University, Moscow, Russia}                   
\centerline{$^{18}$Institute for High Energy Physics, Protvino, Russia}       
\centerline{$^{19}$Lancaster University, Lancaster, United Kingdom}           
\centerline{$^{20}$University of Arizona, Tucson, Arizona 85721}              
\centerline{$^{21}$Lawrence Berkeley National Laboratory and University of    
                   California, Berkeley, California 94720}                    
\centerline{$^{22}$University of California, Davis, California 95616}         
\centerline{$^{23}$University of California, Irvine, California 92697}        
\centerline{$^{24}$University of California, Riverside, California 92521}     
\centerline{$^{25}$Florida State University, Tallahassee, Florida 32306}      
\centerline{$^{26}$University of Hawaii, Honolulu, Hawaii 96822}              
\centerline{$^{27}$Fermi National Accelerator Laboratory, Batavia,            
                   Illinois 60510}                                            
\centerline{$^{28}$University of Illinois at Chicago, Chicago,                
                   Illinois 60607}                                            
\centerline{$^{29}$Northern Illinois University, DeKalb, Illinois 60115}      
\centerline{$^{30}$Northwestern University, Evanston, Illinois 60208}         
\centerline{$^{31}$Indiana University, Bloomington, Indiana 47405}            
\centerline{$^{32}$University of Notre Dame, Notre Dame, Indiana 46556}       
\centerline{$^{33}$Purdue University, West Lafayette, Indiana 47907}          
\centerline{$^{34}$Iowa State University, Ames, Iowa 50011}                   
\centerline{$^{35}$University of Kansas, Lawrence, Kansas 66045}              
\centerline{$^{36}$Kansas State University, Manhattan, Kansas 66506}          
\centerline{$^{37}$Louisiana Tech University, Ruston, Louisiana 71272}        
\centerline{$^{38}$University of Maryland, College Park, Maryland 20742}      
\centerline{$^{39}$Boston University, Boston, Massachusetts 02215}            
\centerline{$^{40}$Northeastern University, Boston, Massachusetts 02115}      
\centerline{$^{41}$University of Michigan, Ann Arbor, Michigan 48109}         
\centerline{$^{42}$Michigan State University, East Lansing, Michigan 48824}   
\centerline{$^{43}$University of Nebraska, Lincoln, Nebraska 68588}           
\centerline{$^{44}$Columbia University, New York, New York 10027}             
\centerline{$^{45}$New York University, New York, New York 10003}             
\centerline{$^{46}$University of Rochester, Rochester, New York 14627}        
\centerline{$^{47}$State University of New York, Stony Brook,                 
                   New York 11794}                                            
\centerline{$^{48}$Brookhaven National Laboratory, Upton, New York 11973}     
\centerline{$^{49}$Langston University, Langston, Oklahoma 73050}             
\centerline{$^{50}$University of Oklahoma, Norman, Oklahoma 73019}            
\centerline{$^{51}$Brown University, Providence, Rhode Island 02912}          
\centerline{$^{52}$University of Texas, Arlington, Texas 76019}               
\centerline{$^{53}$Texas A\&M University, College Station, Texas 77843}       
\centerline{$^{54}$Rice University, Houston, Texas 77005}                     
}                                                                             
%
%%% end of author/institution list
%
\maketitle
\begin{abstract}

  We report on a search for second generation leptoquarks (LQ) produced in 
  $p\overline{p}$ collisions at \mbox{$\sqrt{s} = 1.8$~TeV} using the D\O\ 
  detector at Fermilab.  Second generation leptoquarks are assumed to be 
  produced in pairs and to decay to either $\mu$ or $\nu$ and either a strange 
  or a charm quark ($q$).  Limits are placed on $\sigma(p \overline{p} \to LQ 
  \overline{LQ} \to \mu\nu + jets)$ as a function of the mass of the 
  leptoquark.  For equal branching ratios to $\mu q$ and $\nu q$,  
  second generation scalar leptoquarks with a mass below 160~GeV/$c^2$, vector 
  leptoquarks with anomalous minimal vector couplings with a mass below 
  240~GeV/$c^2$, and vector leptoquarks with Yang-Mills couplings with a 
  mass below 290~GeV/$c^2$, are excluded at the 95\%\ confidence level.
\end{abstract}
\twocolumn
%
%%%%%%%%%%%%%%%%%% MAIN TEXT %%%%%%%%%%%%%%%%%%%%%%%%%%%%%%%%%%%%%%%%%%%%%%%
%
%%%%%%%%%%%%%%%%%% Introduction %%%%%%%%%%%%%%%%%%%%%%%%%%%%%%%%%%%%%%%%%%%%
%
  Leptoquarks (LQ) are hypothetical particles that carry color,
  fractional electric charge, and both lepton and baryon number.  
  They appear in several extended gauge theories and composite models 
  beyond the standard model\cite{generic_lq}.  Leptoquarks with universal 
  couplings to all lepton flavors would give rise to flavor-changing 
  neutral currents, and are therefore tightly constrained by experimental 
  data\cite{fcnc}.  To satisfy experimental constraints on flavor-changing 
  neutral currents, leptoquarks that couple only to second generation 
  leptons and quarks are considered.{\par}

  This Letter reports on a search for second generation leptoquark pairs
  produced in $p\overline{p}$ interactions at a center-of-mass energy 
  $\sqrt{s}$ = 1.8 TeV.  They are assumed\cite{pair_pro} to be produced 
  dominantly via the strong interaction, \mbox{$p\overline{p} \rightarrow g + X 
  \rightarrow LQ\overline{LQ} + X$}.  The search is conducted for the signature 
  where one of the leptoquarks decays via \mbox{LQ $\rightarrow$ muon + quark} 
  and the other via \mbox{LQ $\rightarrow$ neutrino + quark}, where the quark 
  may be either a strange or a charm quark. The corresponding experimental 
  cross section is \mbox{$2\beta(1 - \beta) \times \sigma(p\overline{p} 
  \rightarrow LQ \overline{LQ})$} with $\beta$ the unknown branching fraction 
  to a charged lepton ($e,\mu,\tau$) and a quark (jet) and ($1 - \beta$) the 
  branching fraction to a \mbox{neutrino ($\nu$)} and a jet.  The search 
  considers leptoquarks with scalar or vector couplings in the $\mu \nu + jets$ 
  final state.  Additional details on this analysis may be found in reference 
  \citen{mywork}.  Previous studies by the D\O\cite{D02GenLQ,1GenVLQ} and 
  CDF\cite{CDF2GenLQ} collaborations have considered the $\mu \mu + jets$ 
  final state for scalar couplings, resulting in limits of 140~GeV/$c^2$ and 
  160~GeV/$c^2$ respectively for \mbox{$\beta$ = 1/2.}{\par}

  The D\O\ detector\cite{d0nim} consists of three major components: an inner
  detector for tracking charged particles, a uranium--liquid argon calorimeter
  for measuring electromagnetic and hadronic showers, and a muon spectrometer
  consisting of a magnetized iron toroid and three layers of drift tubes.  Jets 
  are measured with an energy resolution of approximately $\sigma(E)$ = 
  0.8/$\sqrt{E}$ ($E$ in GeV).  Muons are measured with a momentum resolution 
  $\sigma(1/p) = 0.18(p-2)/p^2 \oplus 0.003$ ($p$ in GeV/$c$).{\par}
  
  Event samples are obtained from triggers requiring the presence of a muon
  candidate with transverse momentum \mbox{$p_T^{\mu} >$ 5~GeV/$c$} in the fiducial
  region $|\eta_{\mu}| < 1.7$ (\mbox{$\eta \equiv -\ln[\tan(\frac{1}{2}\theta)]$}, 
  where $\theta$ is the polar angle of the track with respect to the $z$ axis 
  taken along the proton beam line), and at least one jet candidate with 
  transverse energy $E_T^j >$ 8~GeV and $|\eta_j| <$ 2.5.  The data 
  used for this analysis correspond to an integrated luminosity of 94$\pm$5 
  pb$^{-1}$ collected during the 1993--1995 and 1996 Tevatron collider runs at 
  Fermilab.{\par}

  In the final event sample, muon candidates are required to have a reconstructed 
  track originating from the interaction region consistent with a muon of 
  \mbox{$p_T^{\mu} >$ 25~GeV/$c$} and $|\eta_{\mu}| < 0.95$.  To reduce backgrounds 
  from heavy quark production, muons must be isolated from jets ($\Delta {\cal R}
  (\mu,jet) > 0.5$ for $E_T^j >$ 15~GeV, where $\Delta {\cal R}(\mu,jet)$ is the 
  separation between the muon and jet in the $\eta - \phi$ plane), and have energy 
  deposition in the calorimeter consistent with that of a minimum ionizing 
  particle.  Events are required to have one muon satisfying these requirements.  
  Events containing a second muon which satisfy these requirements, with the
  fiducial requirement relaxed to $|\eta_{\mu}| < 1.7$, are rejected.
  {\par}
  
  Jets are measured in the calorimeters and are reconstructed 
  using a cone algorithm with a radius ${\cal R} = 0.5$ $({\cal R} 
  \equiv \sqrt{\Delta\phi^2 + \Delta\eta^2})$.  Jets must be produced within 
  $|\eta_j|<2.0$, and have $E_T^j > 15$~GeV; with the most energetic jet 
  in each event required to have $|\eta_j|<1.5$.{\par}
  
  The transverse energy of the neutrino is not directly measured, but is 
  inferred from the energy imbalance in the calorimeters 
  and the momentum of the reconstructed muon.  Events are required to have 
  missing transverse energy 
  ${\mbox{${\hbox{$E$\kern-0.6em\lower-.1ex\hbox{/}}}_T$}}>30$~GeV. To ensure that 
  $\mbox{${\hbox{$E$\kern-0.6em\lower-.1ex\hbox{/}}}_T$}$ is not 
  dominated by mismeasurement of the muon $p_T$, events having 
  $\mbox{${\hbox{$E$\kern-0.6em\lower-.1ex\hbox{/}}}_T$}$ 
  within $\pi \pm 0.1$ radians of the muon track in azimuth are rejected.{\par}

  To provide further rejection against dimuon events in which one of the muons 
  was not identified in the spectrometer, muons are identified 
  by a pattern of isolated energy deposited in the longitudinal segments 
  of the hadronic calorimeter \cite{zgamma}.  Any event where such  
  deposited energy lies along a track originating from the interaction 
  vertex in the region $|\eta| < 1.7$ and is within 0.25 radians in azimuth 
  of the direction of the 
  {\mbox{${\hbox{$E$\kern-0.6em\lower-.1ex\hbox{/}}}_T$}} vector is rejected.{\par}
  
  Each candidate event is required to pass a selection based on the expected LQ 
  event topology. Since the decay products of the LQ are $\mu q$ or 
  $\nu q$, the muon and neutrino in LQ pair decays come from different 
  parent particles nearly at rest  and are therefore uncorrelated.  For the
  primary background events (e.g. $W + jets$), the two leptons have the same 
  parent.  Similar reasoning holds for the jets.  Correlated backgrounds are 
  rejected with the requirement of significant separation between the muon and 
  $\mbox{${\hbox{$E$\kern-0.6em\lower-.1ex\hbox{/}}}_T$}$ ($|\Delta \phi(\mu,
  \mbox{${\hbox{$E$\kern-0.6em\lower-.1ex\hbox{/}}}_T$})| > 0.3$) and between the 
  two leading jets ($\Delta {\cal R}(j_1,j_2) > 1.4$).{\par}
%
%%%%%%%%%%%%%%%%%%%%%%%%%%%%% Monte Carlo %%%%%%%%%%%%%%%%%%%%%%%%%%%%%%%%%%%%
%
  The {\footnotesize{ISAJET}} \cite{isajet} Monte Carlo event generator is used
  to simulate the scalar leptoquark ($S_{LQ}$) signal, and 
  {\footnotesize{PYTHIA}}\cite{pythia} is used for the vector leptoquark ($V_{LQ}$) 
  signal.  The efficiencies for $V_{LQ}$ and $S_{LQ}$ are consistent 
  within differences due to the choice of generator.  This is verified by 
  choosing a test point at which both scalar and vector Monte Carlo events 
  from the same generator are compared.  Therefore, efficiencies obtained from 
  the two simulations are not distinguished. In addition, the 
  efficiencies for vector leptoquarks are insensitive to differences between 
  minimal vector ($\kappa_G$ = 1; $\lambda_G$ = 0\cite{VLQCouple}) and Yang-Mills 
  ($\kappa_G$ = 0; $\lambda_G$ = 0\cite{VLQCouple}) couplings at large mass 
  \cite{1GenVLQ} \mbox{($M_{V_{LQ}} >$ 200 GeV/$c^2$)}.  The leptoquark 
  production cross sections used for the $S_{LQ}$ are from next-to-leading order 
  (NLO) calculations\cite{kraemer} with a renormalization 
  scale $\mu = M_{S_{LQ}}$ and uncertainties determined from variation of the
  renormalization/factorization scales from $2 M_{S_{LQ}}$ to $\frac{1}{2} 
  M_{S_{LQ}}$.  The $V_{LQ}$ cross sections are leading order (LO) 
  calculations at a scale $\mu = M_{V_{LQ}}$\cite{VLQCouple}.{\par}

  The dominant backgrounds, from  $W+jets$ and \mbox{$Z+jets$}, are simulated using 
  {\footnotesize{VECBOS}}\cite{vecbos} for parton level generation and 
  {\footnotesize{HERWIG}} \cite{herwig} for parton fragmentation.  Background 
  due to $WW$ production is simulated with {\footnotesize{PYTHIA}}\cite{pythia}.
  Additional background from $t\overline{t}$ decays into one or more muons and two 
  or more jets, is simulated using the {\footnotesize{HERWIG}} Monte Carlo program 
  for a top  quark mass of 170~GeV/$c^2$.  Monte Carlo samples are processed 
  through a detector simulation program based on the {\footnotesize{GEANT}}
  \cite{geant} package.{\par}
%
%%%%%%%%%%%%%%%%%%%%%%%%%%% Mu-Nu jj Search %%%%%%%%%%%%%%%%%%%%%%%%%%%%%%%%%%%
%
  With the initial data selection described above, there are 107 events, 
  consistent with a background of $106{\pm}30$ events
  (see Fig. 1).  The dominant background is $W + jets$ with 100$\pm$30 events. 
  Other backgrounds are 2.7$\pm$0.7 ($Z+jets$), 2.4$\pm$0.8 $t\overline{t}$, and
  1.5$\pm$0.6 ($WW$).  The uncertainty in the background is dominated by 
  the statistical uncertainty in the  $W + jets$ simulation and the systematic 
  uncertainty in the $W + jets$ cross section.  The expected signal for 
  160~GeV/$c^2$ scalar leptoquarks is $4.8{\pm}0.7$ events. Signal estimations are 
  shown for a $S_{LQ}$ mass of 160~GeV/$c^2$ using the NLO cross section with a 
  scale of $2 M_{S_{LQ}}$.{\par}

  \begin{figure}
  \vbox{
  \vspace{-0.40in}
  \centerline{
  \psfig{figure=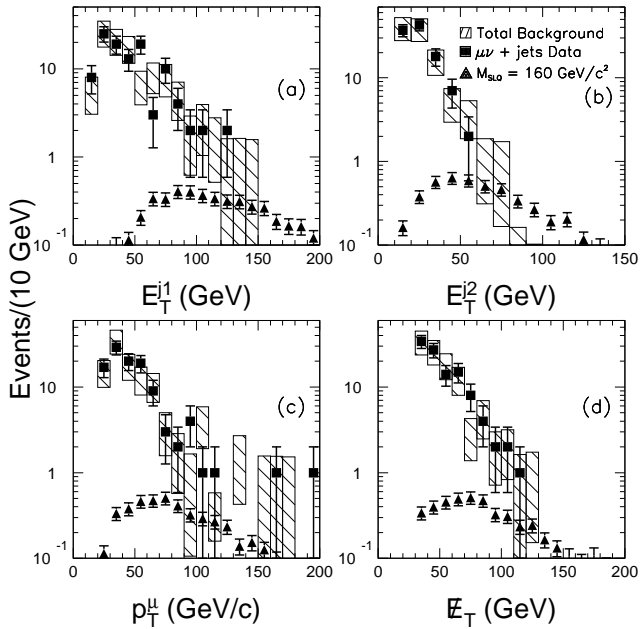,width=3.5in}}
  \vspace{0.05in}
  \caption{Kinematic distributions for $\mu \nu+jets$ events.  The quantities 
  shown in (a)--(d) are used as inputs into the neural network (see text).  The 
  shaded regions give the background expectations, the square points are the 
  $\mu\nu + jets$ data, and the triangular points are signal Monte Carlo.}
  \label{fig:fig1}
  }
  \end{figure}

  \begin{figure}
  \vbox{
  \vspace{-0.40in}
  \centerline{
  \psfig{figure=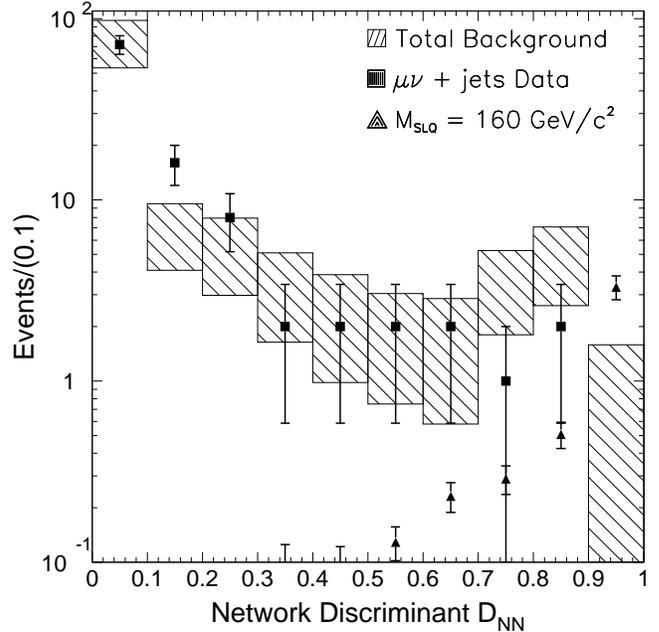,width=3.5in}}
  \vspace{0.05in}
  \caption{Output of the neural network.  The network calculates a value for each 
  event based on the inputs (shown in Fig. 1) and a set of internal values which 
  are determined during network training on signal and background Monte Carlo.}
  \label{fig:fig2}
  }
  \end{figure}
  
  To separate any possible signal from the backgrounds, a neural 
  network (NN)\cite{nn} with inputs: $E_T^{j_1},E_T^{j_2},p_T^{\mu}$ and 
  $\mbox{${\hbox{$E$\kern-0.6em\lower-.1ex\hbox{/}}}_T$}$ 
  and nine nodes in a single hidden layer is used.  The network is trained on a 
  mixture of $W + jets$, $Z + jets$ and $t\overline{t}$ background Monte Carlo 
  events, and an independently generated signal Monte Carlo sample at a mass of 
  160~GeV/$c^2$.  Figure 1 shows distributions of the four input quantities  and 
  Fig. 2 the network output (referred to as the discriminant, $D_{\text{NN}}$).  No 
  evidence of a signal is observed in either the discriminant distribution or any 
  of the kinematic distributions.  For setting limits, the selection on 
  $D_{\text{NN}}$ is optimized by maximizing a measure of 
  sensitivity\cite{sensitivity} defined by

  \begin{displaymath}
    S(D_{\text{NN}}) \equiv\sum_{k=0}^{n}P(k,b)M^{95\%}_A(k,b,s(M_{LQ}))
  \end{displaymath}
  
  \hspace{-0.2in} where $P(k,b)=e^{-b}b^{k}/k!$ is a Poisson coefficient with 
  $k$ being any possible number of observable events, $b$ the expected mean number 
  of background events, and $s(M_{LQ})$ the expected signal for a given leptoquark 
  mass. $M^{95\%}_A$ is an approximate\cite{mapprox} mass limit at the 95\%\ 
  confidence level for a given $k$, $s$ and $b$.  $S(D_{\text{NN}})$ is the sum of 
  the approximate mass limits, weighted by the probability of observing $k = 
  0,1,2,~{\ldots}~,n$($P(n,b)<0.05$) events for a particular choice of the 
  $D_{\text{NN}}$ selection criterion.{\par}
  
  \begin{table}[t]
  \begin{tabular} { c c c c c c }
    $LQ$ Mass  & efficiency & 
    ${\sigma}^{95\%}$ & 
    {\scriptsize BR}$\times$${\sigma}_{S_{LQ}}$ & 
    {\scriptsize BR}$\times$${\sigma}_{MV}     $ &
    {\scriptsize BR}$\times$${\sigma}_{YM}     $  \\
    (GeV/$c^2$) & (\%) & (pb) & (pb) & (pb) & (pb) \\
  \hline

  $100$  & $3.7 {\pm}0.2{\pm}0.6$ &  $0.94$  &  $2.8 $  &  $53  $  &  $430 $  \\
  $120$  & $5.0 {\pm}0.2{\pm}0.7$ &  $0.72$  &  $2.2 $  &  $23  $  &  $150 $  \\
  $140$  & $7.2 {\pm}0.3{\pm}1.1$ &  $0.47$  &  $0.75$  &  $10  $  &  $50  $  \\
  $160$  & $10.3{\pm}0.3{\pm}1.5$ &  $0.33$  &  $0.34$  &  $4.0 $  &  $25  $  \\
  $180$  & $12.2{\pm}0.3{\pm}1.8$ &  $0.27$  &  $0.16$  &  $2.0 $  &  $10  $  \\
  $200$  & $13.4{\pm}0.3{\pm}2.0$ &  $0.25$  &  $0.08$  &  $1.0 $  &  $5.0 $  \\
  $220$  & $14.1{\pm}0.3{\pm}2.1$ &  $0.24$  &  $0.04$  &  $0.45$  &  $2.5 $  \\
  $240$  & $15.2{\pm}0.3{\pm}2.3$ &  $0.23$  &  $0.02$  &  $0.23$  &  $1.3 $  \\
  $260$  & $15.5{\pm}0.3{\pm}2.3$ &  $0.22$  &  $0.01$  &  $0.13$  &  $0.60$  \\
  $280$  & $16.3{\pm}0.4{\pm}2.4$ &  $0.21$  &  $    $  &  $0.06$  &  $0.30$  \\
  $300$  & $15.7{\pm}0.4{\pm}2.3$ &  $0.22$  &  $    $  &  $0.03$  &  $0.18$  \\
  $350$  & $16.4{\pm}0.4{\pm}2.4$ &  $0.21$  &  $    $  &  $    $  &  $0.03$  \\
  $400$  & $17.2{\pm}0.4{\pm}2.6$ &  $0.20$  &  $    $  &  $    $  &  $    $

  \end{tabular}
  \vspace{0.1in}
  \caption{Signal detection efficiencies (with statistical and systematic 
  uncertainty) and cross section limits (95\%\ CL) for leptoquarks in the 
  ${\mu \nu}+jets$ decay channel.  Also shown for comparison are the expected cross 
  sections for $\beta = \frac{1}{2}$.  ${\sigma}_{S_{LQ}}$ denotes
  the theoretical cross section for scalar leptoquarks with a scale 2$M_{S_{LQ}}$, 
  ${\sigma}_{MV}$ the cross section for vector leptoquarks with anomalous minimal 
  vector couplings, and ${\sigma}_{YM}$ leptoquarks with vector Yang-Mills 
  couplings.}

  \label{tab:table1}
  \end{table}

  By maximizing the value of $S(D_{\text{NN}})$ a discriminant selection of
  $D_{\text{NN}} > 0.9$ is obtained. With this selection, no events remain in the 
  data, which is consistent with an expected background of $0.7{\pm}0.9$ events.  
  The remaining background is dominated by $t\overline{t}$ (0.6$\pm$0.2 events). 
  The uncertainty on the total background is dominated by the statistical and 
  systematic uncertainties from $W + jets$.{\par}
  
  Table I shows the signal detection efficiencies and upper limits\cite{limits} on 
  the cross section at the 95\%\ confidence level as a function of the leptoquark 
  mass.  The dominant systematic uncertainty on the signal efficiency is due to the 
  simulation, (initial and final state radiation, parton distribution function, 
  renormalization scale, choice of generator) with a 10\%\ uncertainty.  The 
  systematic uncertainties shown include approximately equal contributions from 
  uncertainty in the jet energy scale\cite{jetenergy} and the trigger 
  efficiency/spectrometer resolution for high $p_T$ muons (6.6\%\ and 6.4\%\ 
  respectively).  The overall systematic uncertainty for the signal efficiency is 
  15\%.{\par}  
  
  \begin{figure}[t]
  \vbox{
  \vspace{-0.20in}
  \centerline{
  \psfig{figure=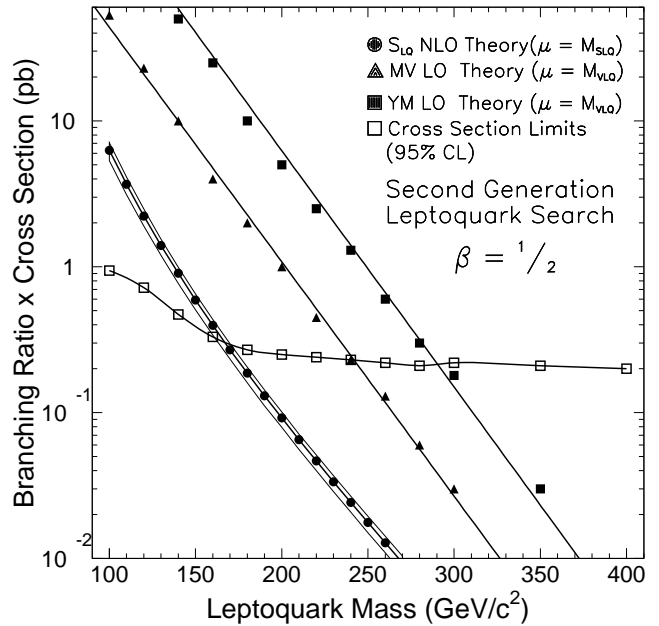,width=3.5in}}
  \vspace{0.05in}
  \caption{Cross section limits in the ${\mu \nu}+ jets$ channel.  The 
  $V_{LQ}$ cross sections are leading order[12], calculated at a scale 
  $\mu = M_{V_{LQ}}$.  The $S_{LQ}$ cross sections are next-to-leading order[13].  
  The calculation is done at a renormalization scale $\mu = M_{S_{LQ}}$ with 
  uncertainties obtained from variation of the renormalization/factorization 
  scale from $2 M_{S_{LQ}}$ to $\frac{1}{2}M_{S_{LQ}}$.  For the $S_{LQ}$ the 
  limit is obtained at the intersection of the experimental curve with the 
  theoretical curve for $\mu = 2M_{S_{LQ}}$.}
  \label{fig:fig3}
  }
  \end{figure}

%%%\cite{kraemer}    = NLO cs = ref #13
%%%\cite{VLQCouple}  =  LO cs = ref #12

  The limits on the observed cross section are shown in \mbox{Fig. 3}, and 
  are compared with the theoretical cross section times branching ratio for 
  scalar and vector leptoquark 
  production for $\beta = \frac{1}{2}$.  Mass limits of 160~GeV/$c^2$ for scalar 
  leptoquarks and 290 (240)~GeV/$c^2$ for vector leptoquarks with Yang-Mills 
  (minimal vector) couplings,  are obtained at the 95\%\ confidence level.{\par}
  
  In conclusion, we have performed a search for second generation leptoquarks
  in the $\mu \nu + jets$ decay channel using $94 \pm 5$~pb$^{-1}$ of data 
  collected with the D\O\ detector at the Fermilab Tevatron.  No evidence 
  for a signal is seen and limits are set at the 95\%\ confidence level on the 
  mass of second generation leptoquarks.  For equal branching fractions to 
  $\mu q$ and $\nu q$ ($\beta = \frac{1}{2}$) limits of 160~GeV/$c^2$, 
  240~GeV/$c^2$, and 290~GeV/$c^2$ for $S_{LQ}$, minimal vector, and Yang-Mills 
  vector couplings, respectively, are obtained.{\par}
%
%%%%%%%%%%%%%%%%%%%%%%%%%%% Acknowledgement paragraph %%%%%%%%%%%%%%%%%%%%%%%%%
%
% Acknowledgement_paragraph.tex
%
We thank the Fermilab and collaborating institution staffs for
contributions to this work and acknowledge support from the 
Department of Energy and National Science Foundation (USA),  
Commissariat  \` a L'Energie Atomique (France), 
Ministry for Science and Technology and Ministry for Atomic 
   Energy (Russia),
CAPES and CNPq (Brazil),
Departments of Atomic Energy and Science and Education (India),
Colciencias (Colombia),
CONACyT (Mexico),
Ministry of Education and KOSEF (Korea),
and CONICET and UBACyT (Argentina).
%
%
%%%%%%%%%%%%%%%%%%%%%%%%%%% References %%%%%%%%%%%%%%%%%%%%%%%%%%%%%%%%%%%%%%%%
%

%
%%% End of Document
%

\begin{references}
\bibitem{mapprox}
  The approximation used in $M^{95\%}_A$ is that uncertainties are not included.
  The difference between $S(0.9)$ calculated using $M^{95\%}_A$ and $S(0.9)$ using 
  an exact calculation is approximately 1\%.{\par}
\bibitem{limits}
  Limits are calculated using a Bayesian approach with a flat prior 
  distribution for the signal cross section.  The statistical and systematic 
  uncertainties on the efficiencies, the integrated luminosity, and the background 
  estimation are included in the calculation (assuming Gaussian prior 
  distributions).{\par}
\end{references}
\end{document}